\documentstyle[aps,preprint,epsfig]{revtex}
\tightenlines
\begin{document}
\newcommand{\beq}{\begin{equation}}
\newcommand{\eeq}{\end{equation}}
\newcommand{\ber}{\begin{eqnarray}}
\newcommand{\eer}{\end{eqnarray}}
\newcommand{\berr}{\begin{eqnarray*}}
\newcommand{\eerr}{\end{eqnarray*}}
\newcommand{\bt}{\beta}
\newcommand{\f}{\beta_f}
\newcommand{\ba}{\beta_a}
\newcommand{\bv}{\beta_v}
\newcommand{\bvc}{\beta_{vc}}
\newcommand{\ns}{N_\sigma}
\newcommand{\nt}{N_\tau}
\newcommand{\lf}{\left\langle L_f \right\rangle}
\newcommand{\llf}{\left\langle | L_f | \right\rangle}
\newcommand{\clf}{\chi_{|L_f|}}
\newcommand{\la}{\left\langle L_a \right\rangle}
\newcommand{\pq}{\left\langle P \right\rangle}
\newcommand{\DD}{\mathcal D}
\newcommand{\LL}{\mathcal L}
\newcommand{\sg}{\sigma_p}
\newcommand{\sm}{\sigma_c}
\newcommand{\se}{\sigma_e}
\newcommand{\tf}{{\rm Tr}_f}
\newcommand{\lm}{\lambda}
\newcommand{\gm}{\gamma}
\newcommand{\al}{\alpha}
\draft
\preprint{\hbox{\begin{tabular}{c}
                hep-lat/9909139 \\
		MRI-PHY/P990929 \\
                TIFR/TH/99-50 \\
\end{tabular}}}
\title{$Z_2$ Monopoles, Vortices, and The Deconfinement 
Transition in Mixed Action SU(2) Gauge Theory}
\author{Saumen Datta\footnote{E-mail:saumen@mri.ernet.in}}
\address{The Mehta Research Institute of Mathematics and Mathematical Physics,
Chhatnag Road, Jhusi, Allahabad 211019, India}
\author{R. V. Gavai\footnote{E-mail:gavai@theory.tifr.res.in}}
\address{Department of Theoretical Physics,Tata Institute of Fundamental 
Research, Homi Bhabha Road, Mumbai 400005, India}
\maketitle
\begin{abstract}
Adding separate chemical potentials $\lambda$ and $\gamma$ for
$Z_2$-monopoles and vortices respectively in the Villain form of the
mixed fundamental-adjoint action for the SU(2) lattice gauge theory,
we investigate their role in the interplay between the deconfinement and
bulk phase transitions using Monte Carlo techniques.  Setting $\lambda$
to be nonzero, we find that the line of deconfinement transitions is
shifted in the coupling plane but it behaves curiously also like the
bulk transition line for large enough adjoint coupling, as for 
$\lambda=0$.  In a narrow range of couplings, however, we find 
separate deconfinement and bulk phase transitions on the same lattice
for nonzero and large $\lambda$, suggesting the two to be indeed 
coincident in the region where a first order deconfinement phase 
transition is seen.  In the limit of large $\lambda$ and $\gamma$, 
we obtain only lines of second order deconfinement phase transitions,
as expected from universality. 
\end{abstract}

\section{INTRODUCTION}
\label{sc.introduction}

The continuum limit of a lattice regularized gauge theory is defined
at its critical point where the lattice correlation length is infinite.
One therefore expects that a large number of lattice actions, differing 
from each other by only irrelevant terms in the sense of the renormalization 
group, describe the same continuum physics.

For pure gauge theories, the simplest discretized action, the Wilson
action \cite{wilson}, is widely used in both analytical and numerical
studies.  It has been very successful in revealing several nonperturbative 
features of gauge theories.  Most such investigations are, however, necessarily
carried out for a finite value of the lattice spacing $a$.  It seems
therefore imperative that the universality of these results is verified
by employing other forms of lattice actions.  Such a study of the
universality of the deconfinement transition for SU(2) gauge theory was
initiated in Ref. \cite{gavai} for the Bhanot-Creutz action
\cite{bhanot} 
\beq
S_{BC} = \f \sum_p \left( 1 - {1 \over 2} {\rm Tr}_f U_p \right)
 + \ba \sum_p \left( 1 - {1 \over 3} {\rm Tr}_a U_p \right)~,
\label{eq.bc}\eeq
where the summation runs over all the plaquettes of the lattice and the
subscript, $f$($a$) indicates that the trace is taken in the fundamental
(adjoint) representation.  The Wilson action corresponds to setting $\ba
= 0$ in Eq. (\ref{eq.bc}). It has a second order deconfinement 
phase transition at which the order parameter $\llf$
acquires a nonzero value, where
\beq
L_f( \vec x) = {1\over 2} {\rm Tr}_f \prod_{\tau=1}^{N_\tau} U_\tau(\vec
x, \tau)~,~~
\label{eq.ldef}\eeq
and $U_\tau$ are the gauge fields in the time direction.  The finite
size scaling analysis \cite{engels} of its susceptibility, 
$\chi_{|L_f|} = \langle L_f^2 \rangle - \llf ^2$, yielded an exponent 
$\omega = 1.93 \pm 0.03$, in good agreement with the corresponding  
value ($1.965 \pm 0.005$) for the three dimensional Ising model.
Universality of the continuum limit predicts a similar deconfinement
transition belonging to the same universality class as the three
dimensional Ising model for all values of $\ba$.

Monte Carlo simulations for Eq. (\ref{eq.bc}) on $ \ns^3 \times \nt $ 
lattices with $\nt = 4$ showed \cite{gavai} that while the second order
deconfinement transition point for the $\ba = 0$ Wilson action entered
the $\f - \ba$ plane as a line of second order transitions, the
transition, surprisingly, turned first order for large enough $\ba$.
This was evident from the facts that i) the order parameter, 
$\llf$, became nonzero discontinuously at the
transition and ii) the exponent of its susceptibility changed
from the Ising model value to 3. If the change of the
order of the deconfinement transition were to persist in the continuum
limit, it would be a serious violation of universality. On the other
hand, the deconfinement line was also found to coincide with the known bulk
transition line\cite{bhanot} for $\nt$ = 4. Studies with varying $\nt$ further
revealed that it scarcely moves as one increases $\nt$, especially in
the region where a strong first order deconfinement transition is
observed\cite{manu}.  

Similar results were also obtained\cite{stephenson} in studies with 
a Villain form of the mixed fundamental-adjoint action\cite{caneschi}, 
\beq
S = \f \sum_p \left(1-{1 \over 2}{\rm Tr}_f U_p\right) + \bv \sum_p
\left(1-{1 \over 2}\sg.{\rm Tr}_f U_p \right),
\label{eq.action}\eeq
where $\sg$ are auxiliary $Z_2$ variables defined on the plaquettes and
the partition function has a sum over all possible values of the $\sg$
variables. This summation ensures that the second term gets
contributions only from the integer representations of SU(2).
The deconfinement phase transition in the $\f$ - $\bv$ plane
of this action was shown\cite{stephenson} to have similar features
as above : the deconfinement transition turned first order for large
enough $\bv$ and the transition line merged with the known bulk
transition line for this action \cite{caneschi}. 

Another example of similar interplay of bulk and deconfinement effects was 
shown\cite{srinath,i1} to be the SO(3) lattice gauge theory, i.e, for $\f$ =0
in Eqs. (\ref{eq.bc}) and (\ref{eq.action}). Being in the same universality
class, it should also have a second order deconfinement phase
transition, although it does not have an order parameter like the
SU(2) theory.  The only transition found in Refs.  \cite{srinath,i1}
for different $\nt$ coincided with the known first order bulk transition.

One way the above conflicting indications can
be interpreted is to postulate that the bulk transition interferes with
the deconfinement line in such a manner as to make the two lines coincident
for a substantial range of $\nt$ and couplings for all the above
actions.  A possible way to test this hypothesis is to modify these
actions further by such additional irrelevant term(s) that the bulk
transitions are either shifted in the coupling plane or entirely eliminated.
The bulk transition in the Villain form of the SO(3) gauge theory is known 
to be caused by a condensation of $Z_2$ monopoles in the strong coupling phase
\cite{halliday}. These monopoles are absent in the weak coupling
region, and can be suppressed at stronger couplings by adding an
irrelevant term to the action.  Investigations \cite{i2} of the monopole
suppressed theory at finite temperature found a continuous
deconfinement transition, very similar to the transition for the SU(2)
theory with Wilson action. This suggests trying an extension of the 
technique of suppression of bulk transitions to the mixed actions as well.  
For the action in Eq. (\ref{eq.action}), the bulk transitions are caused by 
the $Z_2$ monopoles and vortices \cite{caneschi}. In this work, we study its 
finite temperature phase diagram with suppression of these objects.

The plan of our paper is as follows. In the next section, we review
briefly the bulk phase diagram for the action (\ref{eq.action}) and the
results of studies of deconfinement transition for it. Following the SO(3)
results, we suppressed only the $Z_2$ monopoles for it and studied
the resulting theory in the $\f$ - $\bv$ plane. As seen in Sec.
\ref{sc.nomono}, this resulted in suppression of the bulk transition
line coming from the SO(3) axis, but it was found inadequate for
suppressing the remaining bulk transition line, where again the same
interplay of bulk and deconfinement effects was seen.  Nevertheless, we
were able to demonstrate that  a second order deconfinement phase
transition and a first order bulk transition {\it both} exist on the
{\it same} lattice but at different locations in a small range of $\bv$.
In Sec. \ref{sc.nobulk}, we added an additional term to suppress the
$Z_2$ vortices. This resulted in a complete elimination of all the bulk
lines and a universal deconfinement transition in the entire $\f$ -
$\bv$ plane was found. A summary of our results and our conclusions are
presented in Sec. \ref{sc.summary}.

\section{Phase Diagram for Mixed Villain Action}
\label{sc.review}

The phase diagram of the mixed Villain action, Eq. (\ref{eq.action}), was 
studied in Ref. \cite{caneschi} on symmetric lattices and was found to 
be qualitatively similar to that for the Bhanot-Creutz action, 
Eq. (\ref{eq.bc}). As shown in Fig. \ref{fg.review}, lines of first
order transition, emanating from the $\bv \to \infty$ region and the $\f
= 0$ axis, join, and extend to the low $\bv$ phase, before ending at
$\bv \approx 2.2$. The bulk transitions in this action can be
understood to be due to certain $Z_2$ topological objects.
Defining the monopole density, M, and the electric current
density, E, as 
\ber
M &=& 1 - \left\langle {1 \over N_c} \sum_c \sm \right\rangle ~~~{\rm with}~~
\sm = \prod_{p \in \partial c} \sg \nonumber \\
{\rm and}~~ E &=& 1 - \left\langle {1 \over N_l} \sum_l \se \right\rangle 
~~~{\rm with}~~
\se = \prod_{p \in \hat \partial l} \sg \label{eq.objects}
\eer
where $N_c$ and $N_l$ are the number of elementary cubes and links of
the lattice, respectively, the bulk phases are characterized by
presence of condensates of these objects \cite{caneschi}.
The strong coupling phase has both E and M nonzero, corresponding to
condensation of both $Z_2$ electric vortices and magnetic monopoles, while the
large $\f$ and $\bv$ phase, relevant for the continuum
limit, is free of both condensates, as summarized in
Fig. \ref{fg.review} ($\sim 0$ in the figure means zero 
up to $O( \exp(-\f))$ ).

The finite temperature phase transitions for this theory have been
explored in Ref. \cite{stephenson}. As mentioned in
Sec. \ref{sc.introduction}, the results are similar to those obtained
in Refs. \cite{gavai,manu} for the Bhanot-Creutz action. The second
order finite temperature transition line, coming out of the $\bv$ = 0
axis, turns first order and joins the bulk transition line
at its endpoint \cite{f1}.
The finite temperature transition on the $\bv$ axis was studied in Ref.
\cite{i1}. It was found to be a coincident bulk and deconfinement
transition for $\nt$ = 4, 6 and 8 lattices. Since the bulk transition on
the $\bv$ axis is due to the melting of the monopole condensate present
in the strong coupling phase (Fig. \ref{fg.review}), and since the
monopoles are not present in continuum limit, we studied the
deconfinement transition on this axis after suppressing the
monopoles\cite{i2}. The resulting theory i) had no bulk transitions, 
ii) yielded a deconfinement transition very similar to that found for the 
Wilson action, in agreement with expectations based on universality
and iii) displayed a shift in the transition point with $\nt$.  As a 
natural next step, we turn to a study of the mixed theory (action 
(\ref{eq.action})) with monopole suppression, as described below.

\section{Phase Diagram with Monopole Suppression}
\label{sc.nomono}

Adding a chemical potential term for the monopoles as in
Ref. \cite{i2}, the mixed fundamental-adjoint Villain action becomes
\ber
S(U, \sigma) &=& \f \sum_p \left( 1 - {1 \over 2} \tf U_p \right)
\label{eq.chsmono} \\ &{+}& \bv
\sum_p \left( 1 - {1 \over 2}\sg. \tf U_p \right) + \lm \sum_c \left(
 1 - \sm \right), \nonumber
\eer
where the last summation runs over all the elementary 3-cubes of the
lattice, and $\sm$ is defined in Eq. (\ref{eq.objects}). The
classical continuum limit is left unaffected by the 
additional monopole term and one obtains the same continuum
relation, $4 g^{-2} = \f + \bv$, in that limit. 
For SO(3), $\lm$ = 1 was found to be sufficient to suppress the bulk 
transition. We therefore took $\lm$ = 1 for our simulations in the 
entire $\f$ - $\bv$ plane. Our simulations consisted of a 3-steps
iteration.  First, all the gauge variables were updated using Creutz's
heatbath algorithm. This was followed by a heatbath sweep for the $Z_2$ 
variables. In the third step of our iteration, a fraction of the links
(arbitrarily chosen to be $\frac{1}{4}$) were multiplied by a $Z_2$
element subject to a probability determined by the $\f$ term. This
third step is essential for reducing the otherwise enormous
autocorrelations for large $\lm$ simulations, and generalizes the
similar step used in Ref. \cite{i2} for the SO(3) gauge theory. Measurements 
were made after every such compound iteration.  Using hysteresis runs of 15000 
iterations per point, and monitoring the plaquette variables $P =
\langle {1 \over 2} {\rm Tr}_f U_p \rangle$ and $P_a =\langle {1 \over
2} \sg {\rm Tr}_f U_p \rangle$, we mapped out the phase diagram 
on an $8^3\times4$ lattice.  

First order transitions, with significant discontinuities in 
the average plaquette $P$ for $\bv > \f$ and the ``adjoint'' plaquette
$P_a$ for $\bv < \f$, were observed. The transition points are shown by
filled circles in Fig. \ref{fg.bulkmono}. While the transition line in
the large $\bv$ region is very similar to the $\lm = 0$ case
(Fig. \ref{fg.review}), unlike that case the transition line does not
have an endpoint and divides the $\f - \bv$ plane in two disjoint
parts. Also the transition line coming out of the $\f$ = 0 axis in
Fig. \ref{fg.review} is absent here.  Since it is caused by condensation of 
monopoles, which have been suppressed here, its absence was to be expected. 

To look for the deconfinement transition, we monitored the behavior of
$\llf$. The deconfinement transition point on the Wilson axis was seen
to extend into a line of continuous deconfinement transitions as we
switched on $\bv$. On increasing $\bv$, the deconfinement transition
line was seen to merge with the first order line, and the order
parameter $\llf$ showed a discontinuous jump to a nonzero
value, indicating a first order deconfinement transition for these
points. The dotted (solid) line in Fig. \ref{fg.bulkmono}
shows the line of second (first) order deconfinement transitions. 
The facts that unlike the $\lm$ = 0 case, the 1st order line does not
have an endpoint here, and the line coming from the large $\f$ side is
clearly not a deconfining line before the line coming from the Wilson
axis meets it at $\f \sim 2$, give credence to the hypothesis that
the line in the large $\bv$ region is a coincident bulk and
deconfinement transition line and, as we will see later, allows one 
to actually see two separate transitions on the same lattice. In what 
follows, we discuss first the 1st order transition line and then 
the deconfinement transition line in some more detail. 

\subsection{The 1st Order Transition Line}

The first order transition points, shown in Fig. \ref{fg.bulkmono}, 
are listed in Table \ref{tbl.bulkmono} along with the discontinuities
in various observables and the values of $\llf$ at the transition
point on the low $\f ~(\bv)$ side at fixed $\bv ~(\f)$.
In the $\bv > \f$ region, the transition is
associated with large discontinuities in $P$ and $\llf$.
Also $\llf \sim 0$ till the transition point, indicating that
the transition line signals a deconfinement transition. The
transition line in the large $\bv$ region is very similar to 
the $\lm$ = 0 case. This is expected, since in the large
$\bv$ limit, the monopoles are automatically suppressed(see
Fig. \ref{fg.review}) and the additional monopole suppression term 
does not have much effect. As $\bv \to \infty$, the dominant
contribution  to the path integral comes from those configurations 
which have $\sg . \tf U_p = 2$. Now the plaquette variables $\sg$
can be integrated out, constraining the gauge variables $U_l$ to take
values only in the center group $Z_2$ : $ \sg = {1 \over 2} \tf U_p = 
\prod_{l \in \partial p} \sigma_l$ where $\sigma_l$ are $Z_2$
variables defined on the links and the $U_l$ are frozen to the values
$\sigma_l$. Then $\sm$ = 1 for all cubes $c$ in this limit, 
and therefore the additional monopole suppression term
does not have any effect.  The action reduces to that of the $Z_2$
gauge theory, which has a well known first order transition at 
$\f \approx 0.44$ with a discontinuity in the average plaquette.
As $\bv$ is reduced from $\infty$, the transition line starts shifting
from that of the $\lm$ = 0 case. The transition line shown in
Fig. \ref{fg.bulkmono} can be reproduced reasonably well up to $\bv
\sim 2$ by taking into account only configurations where $U_l$ can 
have small fluctuations around $\sigma_l$: \[ U_l = \sigma_l U_l^\prime 
\quad{\rm where} ~U_l^\prime = 1 + i g A_l ~. \] Perturbatively integrating
over the fluctuations yield, in the leading order,
\[ P_a = {1 \over 2} \langle \sg {\rm Tr}_f U_p \rangle = 1 - {3 \over
4 \f} + O({1 \over \f^2}) ~. \]
Up to this order, the integral over $A_l$ only produces a
renormalization of $\f$, and the transition point changes to
\beq
\f \approx {0.44 \over \left(1 - {3 \over 4 \bv} \right)} ~.
\label{eq.estm2}\eeq
The prediction of Eq. (\ref{eq.estm2}), shown in
Fig. \ref{fg.bulkmono}, matches the Monte Carlo results quite well for
$\bv \gtrsim 2$.

In the large $\f$ region, the transition is associated with a large
discontinuity in $P_a$, as can be seen from Table \ref{tbl.bulkmono}.
Also from the $\llf$ value at the lower side of the transition point
one can see that both the sides of the transition are in deconfined
state. The transition line in this region can be understood from the
known bulk transition in the $Z_2$ gauge-Higgs theory. In the $\f \to
\infty$ limit, the gauge variables are frozen : $\tf U_p = 2 ~~\forall
p$. The action in Eq. (\ref{eq.chsmono}) then reduces to 
\beq
S = \bv \sum_p (1-\sg) + \lm \sum_c (1-\sm)~,
\label{eq.chsz2}\eeq
after dropping an irrelevant constant.  It describes a system with only $Z_2$ 
degrees of freedom $\sg$.  Under a duality transformation \cite{savit}, one 
can rewrite it as the action for a $Z_2$ 
gauge-Higgs system (modulo irrelevant constants): 
\beq
\tilde{S} =  - \tilde{\bv} \sum_p \prod_{l \in \partial p} \gamma_l -
\tilde{\lm} \sum_{i,\mu} s_i \gamma_{i, i + \mu} s_{i+\mu}~.
\label{eq.z2higgs}\eeq
Here $s$ and $\gamma$ are $Z_2$ variables residing on the sites and links of 
the dual lattice respectively, and 
\beq
\begin{array}{r}
\tilde{\bv} \\
\tilde{\lm} \\
\end{array}
\left\}~ = ~{1 \over 2} \ln ~ {\rm coth} ~\right\{ 
\begin{array}{l}
\bv \\
\lm \\
\end{array} ~.
\label{eq.dual1}\eeq
The $Z_2$  gauge-Higgs system is known to have a nontrivial phase diagram 
\cite{shenker}, leading to a nontrivial structure in the $\bv$ - $\lm$
plane at $\f = \infty$ for the mixed Villain action. For $\lm$ = 1, a first 
order transition at $\bv \approx 0.44$ is predicted for $\f = \infty$. For large
but finite $\f$, the gauge variables fluctuate around the frozen value
and yield the leading order estimate for the transition point,
\beq
\bv \approx {0.44 \over  \left(1 -{3 \over 4 \f} \right) }~.
\label{eq.estm1}\eeq
This leading order estimate is seen in Fig. \ref{fg.bulkmono} to work
quite well up to $\f \sim 2$. It is also consistent with Table
\ref{tbl.bulkmono}, which shows the transition to have a nearly
continuous $P$ and an almost constant discontinuity in $P_a$ in the
region of very large $\f$.

While the above effective descriptions of the bulk transitions for $\f \to
\infty$ and $\bv \to \infty$ were different, they are probably related
by a duality transformation. In the $\lm \to \infty$ limit, the theory 
is self-dual \cite{gavai2}. In this limit $\sm$ is constrained to be
1, which can be solved as $\sg = \prod_{l \in p} \sigma_l$, where
$\sigma_l$ are $Z_2$ variables defined on links. Now a transformation
$U_l \to U_l \sigma_l$ interchanges the role of the fundamental and
adjoint terms, leaving the action invariant. Under this duality
transformation, $\f \leftrightarrow \bv$ and $P \leftrightarrow P_a$. 
Even for $\lm$ = 1, the constraint $\sm$ = 1 is approximately true, 
and the self-duality is approximately obeyed. 
The locations of first order transition points in  
Fig. \ref{fg.bulkmono} and the discontinuities in the plaquette
variables $P$ and $P_a$ presented in Table \ref{tbl.bulkmono}, display
such a symmetry around the $\f = \bv$ line. Taken together with the
good agreement of Eqs. (\ref{eq.estm2}) and (\ref{eq.estm1}) with the
Monte Carlo results in Fig. \ref{fg.bulkmono}, it leads one to the
identification of the entire first order transition line traced out by
the data points as a bulk transition line. Since the order parameter for
the deconfinement phase transition, $\llf$, becomes nonzero on the same
line for $\bv \gtrsim 0.74$, one has a coincident deconfinement transition
line as well. To investigate the origin of the deconfining nature of
the first order line for $\bv \gtrsim 0.74$, we next turn to an
exploration of the deconfinement line starting from the small $\f$ side.

\subsection {The 2nd Order Transition Lines}

For $\bv$ = 0, the chemical potential term decouples from the SU(2)
part and we have the usual Ising-like second order deconfinement
transition on this axis, at $\f \approx 2.3$ for $\nt$ = 4 lattices
\cite{engels}. One expects, from continuity, a second order transition
also for small $\bv$, as in all the examples mentioned in
Sec. \ref{sc.introduction}. Hysteresis runs at $\bv$ = 0.3, 0.5 and 0.7
did indeed indicate a continuous deconfinement transition. For a more
quantitative study, we studied the $|L_f|$-susceptibility, $\clf$, on
$\ns^3 \times 4$ lattices at each $\bv$ with $\ns$ = 8, 12 and 16. For
each lattice, a low statistics run was made at the transition point
estimated from hysteresis runs, and the peak of the susceptibility
estimated by extrapolating to nearby couplings using Ferrenberg -
Swendsen methods\cite{ferrenberg}. If the peak was not too far from 
the input $\f$, a longer run was made to generate $10^5$
configurations. Otherwise the procedure was repeated at the fresh 
peak location. The susceptibility curves for the three $\bv$ values are 
shown in Fig. \ref{fg.ftmono}. The critical exponent $\omega$ at each 
$\bv$, obtained from a linear fit to ${\rm ln} ~ (\clf)_{max} = \omega ~
{\rm ln} ~ \ns$, is shown in Table \ref{tbl.ftmono} and is seen to be
in good agreement with the $\bv$ = 0 exponent. The average plaquette
$P$ from these runs was smooth everywhere, and the corresponding 
susceptibility peaks did not sharpen with $\ns$ at all, indicating 
a lack of bulk transition at these points. The transition points, 
shown in Fig. \ref{fg.bulkmono} by triangles, are therefore pure 
finite temperature transitions. 

On increasing $\bv$ further, the transition was found to change
its behavior. Already at $\bv$ = 0.74, the rise of Polyakov
loop from zero was associated with a discontinuity, indicating a first
order deconfinement transition. The first order nature of the
transition at this coupling was ascertained by a finite size scaling
study. The distribution of the plaquette and Polyakov loop variables
at the transition point are shown in Fig. \ref{fg.hist4}. They show
the presence of metastable states, and with increase of spatial
lattice size, the two - peak structure was seen to sharpen further
without any visible movement in the peak positions, indicating
a first order transition in the infinite volume limit. Fig.
\ref{fg.hist4} also clearly shows that the transition is a deconfining
one, since the $L_f$ distribution at the lower state is peaked about
zero. 

On increasing $\bv$ further, the same behavior continued : the
deconfinement transition line, characterized by rise of $\llf$ from
$\sim 0$ to a nonzero value, was associated with a discontinuous
jump. Also the average plaquette displayed a discontinuity at the same
point. In Fig. \ref{fg.bulkmono} the complete deconfinement line is
shown with the dotted part indicating second order and the solid part
indicating first order transitions. 
This line is very much reminiscent of the deconfinement
transition line of Refs. \cite{gavai,stephenson}, and is suggestive
of a merger of the deconfinement transition line with the bulk 
transition line.

A possible litmus test of the coincidence scenario is to see two
separate transitions on the same lattice, especially in the vicinity
where the two transition lines meet and remain coincident thereafter.
From the transition points in Table \ref{tbl.bulkmono} and Fig.
\ref{fg.bulkmono} one can see that at $\bv \sim 0.7$, one may see two
separate transitions for $\nt = 4$ lattices.  Such an expectation
was borne out by an explicit study at $\bv$ = 0.7 on an $8^3\times4$
lattice, the results of which are shown in Figs. \ref{fg.twotr}a and
\ref{fg.twotr}b. The former, a hysteresis run performed from the low
$\f$ end, shows that all observables, $P$, $P_a$ and $\llf$ show a jump
at around $\f \approx 2.21$ whereas the latter shows the order parameter
$\llf$ along with its susceptibility, obtained from the longer run
mentioned above.  It clearly shows a second order deconfinement
transition taking place first at $\f \sim 2.1$, followed by a bulk phase
transition later at $\f \sim 2.2$.

We have carried out a similar exercise for $\nt$ = 6 lattices as well,
with essentially similar results. The deconfinement trajectory now
starts from the Wilson axis at $\f \approx 2.43$ \cite{fingberg}. At
$\bv$ = 0.3 and 0.5 we get a continuous transition, with the
transition point shifted to a slightly higher $\f$ (see
Fig. \ref{fg.bulkmono}), as expected for a physical transition. The
susceptibility curves for $\ns$ = 12, 14 and 18 lattices, shown in
Fig. \ref{fg.ftmono6}, and the critical exponent obtained from them,
shown in Table {\ref{tbl.ftmono}, indicate an Ising - like second 
order transition at these points. On increasing $\bv$ further, the
transition line hits the 1st order line and turns first order, but 
now at a slightly smaller value of $\bv$. Already at $\bv = 0.7$ one 
sees a first order transition, as can be seen form the plaquette and 
Polyakov loop distributions shown in Fig. \ref{fg.hist6}. One thus 
needs to choose a slightly smaller value of $\bv$ for $\nt =6$ in
order to see again two transitions on the same lattice but one 
does see them. The $\lm$ = 1 simulations for the mixed Villain action
thus lend a strong credibility to the hypothesis that the
deconfinement transition line for possibly a large range of
$\nt$ merges with the bulk transition line.  Since the latter branches
out in this case and exhibits no end point, the merger is easy to
observe numerically: the small $\bv$ region has only a deconfinement
transition line and the large $\f$ region has only a bulk transition line,
while they seem to be coincident in the $\bv \gtrsim 0.7$ region.

While the $Z_2$ - monopole suppression thus provides convincing
indications that the unphysical bulk transition line is coincident
with the physical deconfinement transition line for a range of $\nt$, 
and thus provides a reasonable explanation for the latter turning first
order in the $\f$ - $\bv$ plane, it is insufficient to remove all the bulk
transitions from the region of interest and to show a second order
transition similar to the Wilson action case for large $\bv$ values.
We therefore turn our attention in the next section to the other 
$Z_2$ objects causing the remaining bulk transition : the electric 
vortices \cite{caneschi}.

\newpage

\section{Suppression of Electric Vortices \\ and Magnetic Monopoles}
\label{sc.nobulk}

In order to suppress the electric vortices in addition to the magnetic
monopoles, we consider the action
\ber
S(U, \sigma) = &{}&\f \sum_p \left( 1 - {1 \over 2} \tf U_p \right)
\nonumber \\ &{}&+ \bv
\sum_p \left( 1 - {1 \over 2} \sg . \tf U_p \right) \nonumber \\ &{}&
+ \lm \sum_c \left(1 - \sm \right) + \gm \sum_l \left( 1 - \se \right),
\label{eq.nobulk}\eer
with $\se$ defined in Eq. (\ref{eq.objects}). The last term
suppresses the $Z_2$ electric current loops. It should be irrelevant
in the continuum limit, leaving the relation between the continuum
coupling and $\f$ and $\bv$ unchanged. For sufficiently large
$\gm$, one expects that the bulk transition line caused by the
condensation of electric loops will be suppressed.  One can reduce the
number of input parameters by sending $\lm \to \infty$ and using the
the explicit self-dual form of the above action.  The monopole term is 
then absent and the plaquette variables $\sg$ are replaced by products over 
corresponding $Z_2$ - link variables. This form has the
advantage of being more tractable numerically. We employed this 
self-dual form for our investigations at large $\gm$.

We followed the same procedure as in the previous section to monitor 
$\llf$ and the finite size scaling behavior of its susceptibility 
for studying the finite temperature transition. $P$ and $P_a$ were 
utilized to look for the bulk transitions. We used a heat-bath
algorithm for both the gauge and $Z_2$ variables. One sweep consisted
of updating all the gauge variables of the lattice, followed
by updating of all the $Z_2$ variables. We made a number of hysteresis
runs of 15000 iterations per point at different $\bv$ values on a
$8^3\times 4$ lattice. No discontinuous transition was found anywhere
in the $\bv - \f$ plane. The plaquette variables $P$ and $P_a$ were
smooth everywhere, indicating that the additional term has succeeded
in eliminating all bulk transitions. The deconfinement transition
points were estimated from the rise in $\llf$.  
In order to look for the expected shift in the deconfinement
transition point, we repeated the exercise on a $12^3 \times 6$
lattice. Our investigations revealed lines of deconfinement 
transition points, consistent with (see Fig. \ref{fg.dcpln})
\beq
\f + \bv \approx \beta_c^W,
\label{eq.dcpln}\eeq
where $\beta_c^W$ is the deconfinement transition point for the
Wilson action ($\approx 2.30$ for $\nt$ = 4 \cite{engels} and 
$\approx 2.43$ for $\nt$ = 6 \cite{fingberg} lattices).
Rapid but continuous rises of $\llf$, associated with diverging
susceptibilities with increasing spatial volume, were found at the
deconfinement transitions at all the $\bv$ values investigated. 
These properties are similar to the known features of the deconfinement
transition for the Wilson action for SU(2).  

The critical exponent $\omega$ was obtained from a 
finite size scaling analysis of the $L_f$-susceptibility on
$\ns^3 \times 4$ lattices with $\ns$ = 8, 12 and 14 at $\bv$
= 1 and 2. Long simulation runs of $10^5$ thermalized configurations
and the usual spectral density methods were used to obtain the
peak height at the transition point for each lattice. The
susceptibility peaks are shown in Fig. \ref{fg.polsusc4}. For 
$\nt$ = 6, we used $\ns$ = 12, 14 and 18 and investigated the $\bv$ 
values 0.5 and 1.5. The results are shown in Fig. \ref{fg.polsusc6}. 
Values of the critical exponent $\omega$ obtained from these peaks are
listed in Table \ref{tbl.nobulk}, and are in good agreement with the
value for the Wilson action for SU(2). This behavior is, of course,
what one expects from universality arguments. Moreover, the
phase transition line, Eq. (\ref{eq.dcpln}), is also consistent
with the expected continuum limit behavior of this action.

Suppressing the artifacts which cause the (first order) bulk transitions
in the mixed Villain action, namely, the $Z_2$ magnetic
monopoles and vortices defined above, thus yield results consistent with 
expectations from universality, both for the order of the transition as
well as the shift of the transition point with $\nt$.  Perturbatively,
one expects the contribution of these topological objects to be zero in
the continuum limit. It is, however, unclear whether this is indeed so
nonperturbatively as well.  In fact, the earlier results for the
unsuppressed action ($\lm = \gm = 0$) suggest that very large $\nt$ ( $ \gg
16$ ) will be needed to see the deconfinement transition separate out
from the bulk transition in the $\f$ - $\bv$ plane. 

\section{Summary and Conclusions}
\label{sc.summary}

Earlier numerical simulations on asymmetric $\ns^3 \times \nt$ lattices
using mixed actions, defined by Eq. (\ref{eq.bc}) or (\ref{eq.action}),
yielded apparently paradoxical results. The deconfinement phase
transition in these cases was found to become first order for a large
range of the adjoint coupling, although one naively expects these
actions to be in the same universality class as the Wilson action,
which is known to have a second order deconfinement transition.
On the other hand, the transition exhibited very little shift
with the temporal lattice size $\nt$ and the average action
was also discontinuous at the transition.  These results could be
consistent with universality if a curious interplay between the physical
deconfinement transition and the bulk transitions, which are merely
lattice artifacts, existed for some unknown reasons.  The absence of
any significant variations in the locations of the transition or any
change in its nature for large $\bv$ made other options of saving
universality rather implausible.

Since variations with $\nt$, a defining characteristic of the
deconfinement phase transition which is not supposed to affect the bulk
transition, failed to resolve the paradoxes, it
seemed natural to seek a mechanism to shift or eliminate the bulk
transition(s), leaving the universality class intact.
Taking a clue from our earlier studies for the SO(3) gauge
theory, where a similar mixture of bulk and deconfinement effects is
seen, we attempted in this paper elimination of the bulk transition(s) by 
suppressing certain topological objects as a means to study the
deconfinement transition in the mixed theory.

First, we considered the mixed Villain action with an addition of a chemical
potential for the $Z_2$ - magnetic monopoles and studied the phase diagram 
on asymmetric lattices in the $\f$ - $\bv$ plane of couplings.
The monopoles get suppressed with increasing chemical potential $\lm$.
Numerical simulations for $\lm$ = 1 showed an interesting phase diagram 
which was quite different from that of the original theory.
Nevertheless, it still shared the paradoxes mentioned above.  We could,
however, demonstrate that i) a second order deconfinement transition
line emanating from the $\f$ axis meets the bulk transition line at
a finite $\bv$, ii) the bulk transition line has no end point,
and iii) the change of the order of the deconfinement phase
transition occurs as the two lines merge. Moreover, we were
able to show the presence of two phase transitions on the same finite 
lattice in the vicinity of the point of merger : a second order
deconfinement phase transition, indicated by a continuous sharp rise in
the order parameter, followed by a discontinuous bulk transition, where
all observables exhibited an abrupt change, but both sides of the
transition were in a deconfined state.

A further suppression of the $Z_2$ electric vortices enabled us to get
rid of the bulk transitions completely, as there was no trace of any
discontinuity anywhere in the coupling space and the average plaquette
variables were smooth everywhere. A study of deconfinement transition in 
this theory revealed a line of second order deconfinement transitions,
obeying $\f + \bv \sim \beta_c^W$, where $\beta_c^W$ is the
corresponding transition point for the Wilson action. Therefore the
removal of the bulk transition lines by suppressing the
topological objects causing those transitions led to results
fully consistent with universality.  Since the terms added to the
action in the process do not contribute in the naive continuum limit,
one can formally attribute the anomalous behavior of the deconfinement 
transition lines for the action in Eq. (\ref{eq.action}) to the
presence of bulk transitions. Of course, it is natural to expect that the
same thing happens for the action in Eq. (\ref{eq.bc}), too. While the vexed
possibility of violation of universality in these cases is thus finally
eliminated, some nagging questions still remain.  Foremost amongst them
is how rapidly the topological objects discussed here vanish from the
lattice for the original theories as one approaches the continuum limit.  
Simply put, even for the original theories one now expects for sure 
the separation of the two lines of transitions for some $\nt$.
However, it is also clear that it must be at an almost 
astronomically large value.  One knows that this is indeed 
the case for spin models\cite{hasenbusch}, where one needs to work at
a billion times larger correlation lengths to see universal behavior in
the adjoint direction.  One does not expect it to be so, however, 
for a continuous gauge symmetric case as ours. Another interesting and 
related question is about the behavior of the order parameter at the
bulk transition when the two transitions do separate : it will have to 
be continuous and zero through the bulk transition and it will also be 
exponentially small after the deconfinement phase transition for very
large $\nt$ due to the divergences associated with the point source
it represents.

\begin{figure}[htbp]\begin{center}
\narrowtext
\epsfig{height=8cm,width=12cm,file=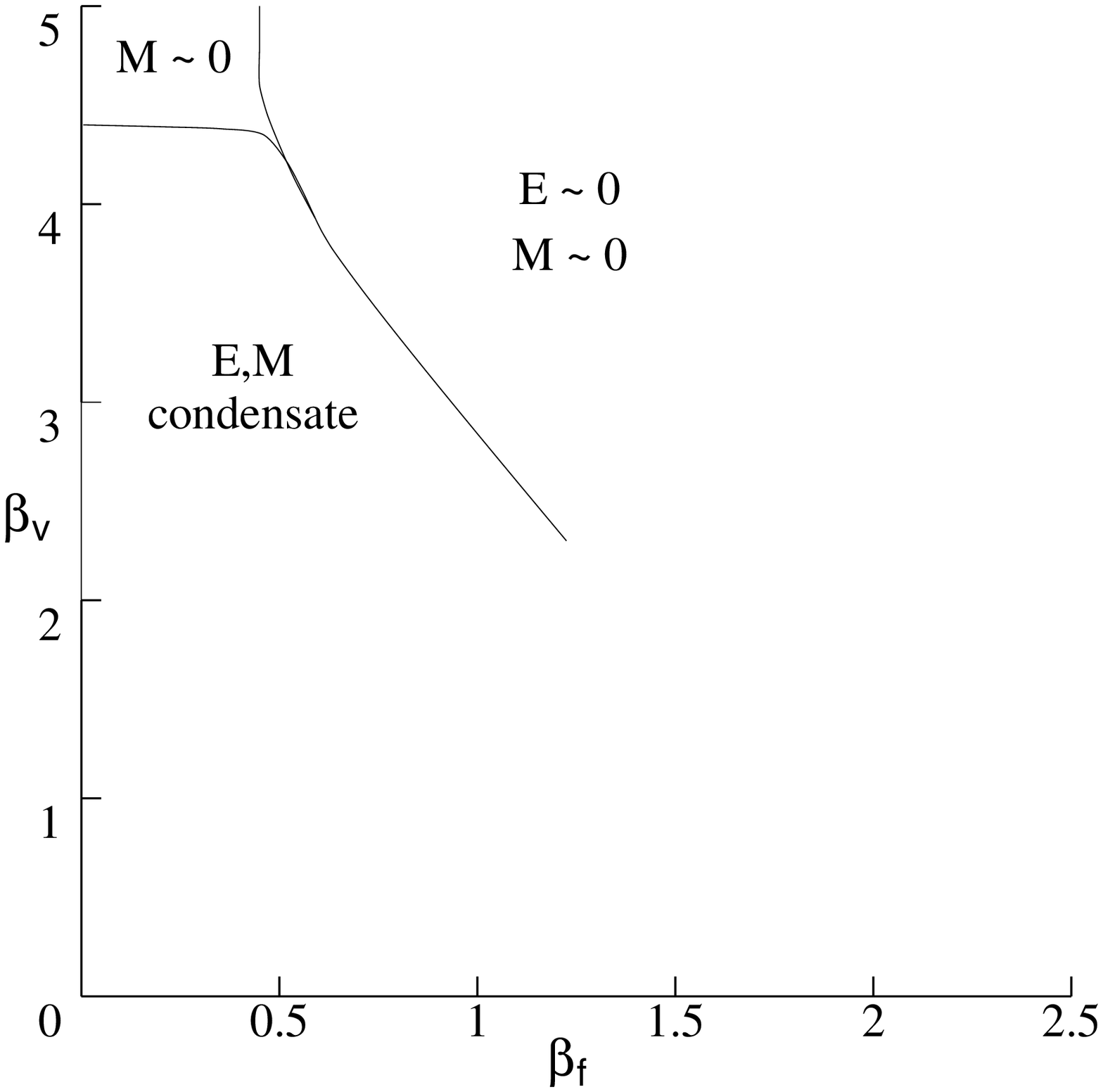}
\caption{The phase diagram for the action (\protect\ref{eq.action}),
showing the first order bulk phase transition lines (from
Ref. \protect\cite{stephenson}). The bulk phases are characterized by values
of E and M, as shown.}
\label{fg.review}\end{center}\end{figure}

\begin{figure}[htbp]\begin{center}
\epsfig{height=8cm,width=12cm,file=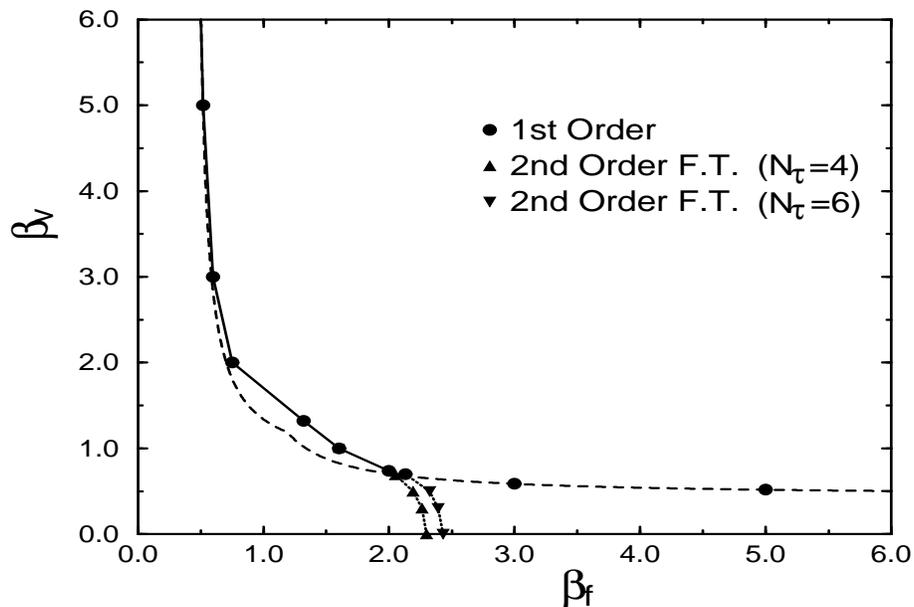}
\caption{The phase diagram for the action (\protect\ref{eq.chsmono})
with $\lm$ = 1 on an $8^3 \times 4$ lattice. The filled circles show first
order transition points. The dashed lines show the estimates from Eqs. 
(\protect\ref{eq.estm2}) and (\protect\ref{eq.estm1}). The triangles 
in the low $\bv$ region show the locations of Ising-like second 
order deconfinement phase transitions on $\nt$ = 4  and 6 lattices 
respectively. The dotted and solid lines show the second and first
order deconfinement transition lines.}
\label{fg.bulkmono}\end{center}\end{figure}

\begin{figure}[htbp]\begin{center}
\epsfig{height=8cm,width=12cm,file=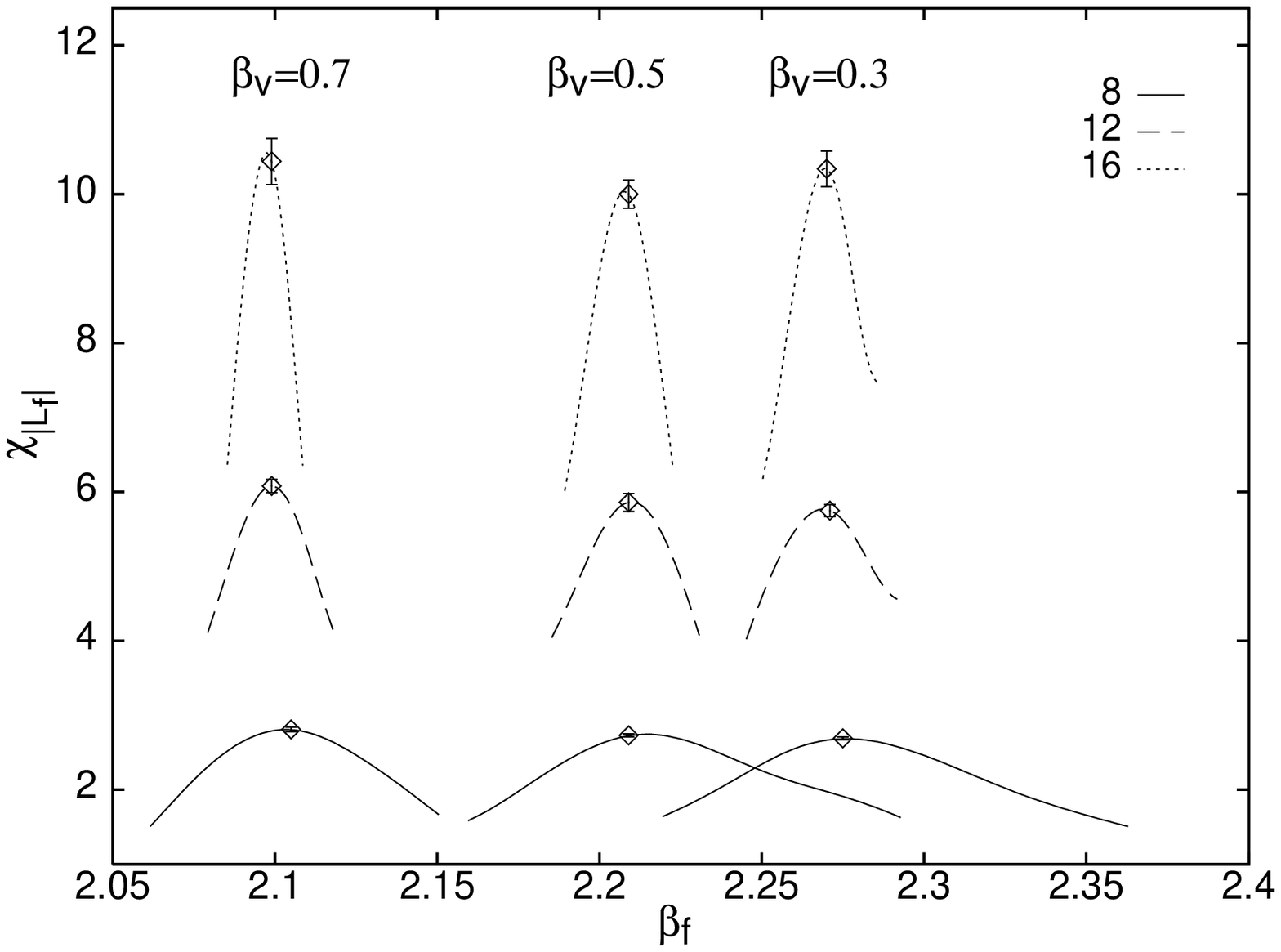}
\caption{$\clf$ as a function of $\f$ for $\nt$ = 4 lattices with
$\ns$ = 8, 12 and 16 for the action (\protect\ref{eq.chsmono}) at 
$\bv$ = 0.7, 0.5 and 0.3.  The values obtained directly from the
long simulation runs are also shown with error bars. 
The lines were obtained using Ferrenberg-Swendsen extrapolation.}
\label{fg.ftmono}\end{center}\end{figure}

\begin{figure}[htbp]\begin{center}
\epsfig{height=8cm,width=12cm,file=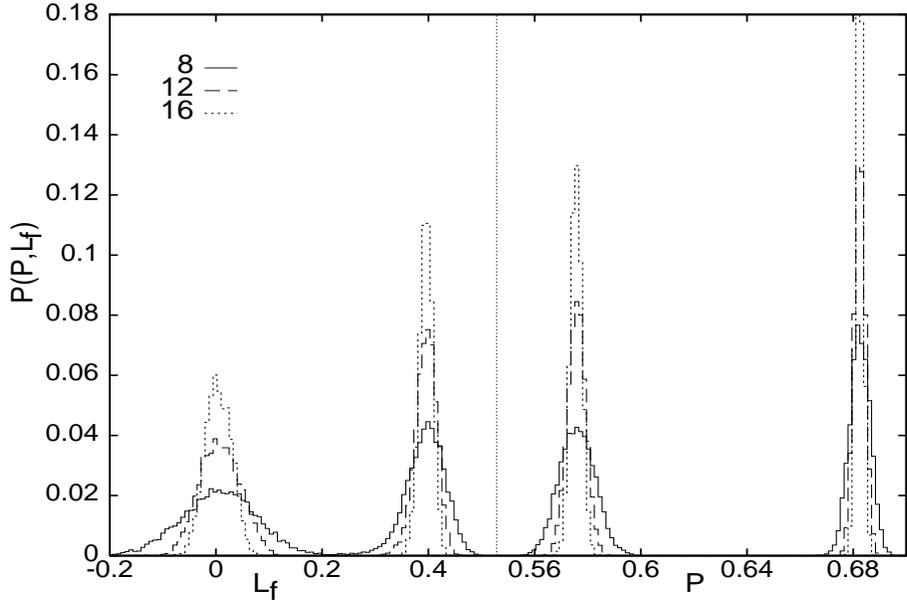}
\caption{Probability distributions of $L_f$ and the plaquette variable
at $\bv$ = 0.74, $\f$ = 2, for $\nt$ = 4 lattices with $\ns$ = 8, 12
and 16. Two peak structures are visible that sharpen with increase in
lattice size, indicating a first order transition. The $L_f$ peaks
clearly indicate that the transition is a deconfining one.}
\label{fg.hist4}\end{center}\end{figure}

\newpage

\begin{figure}[htbp]\begin{center}
\epsfig{height=8cm,width=8cm,file=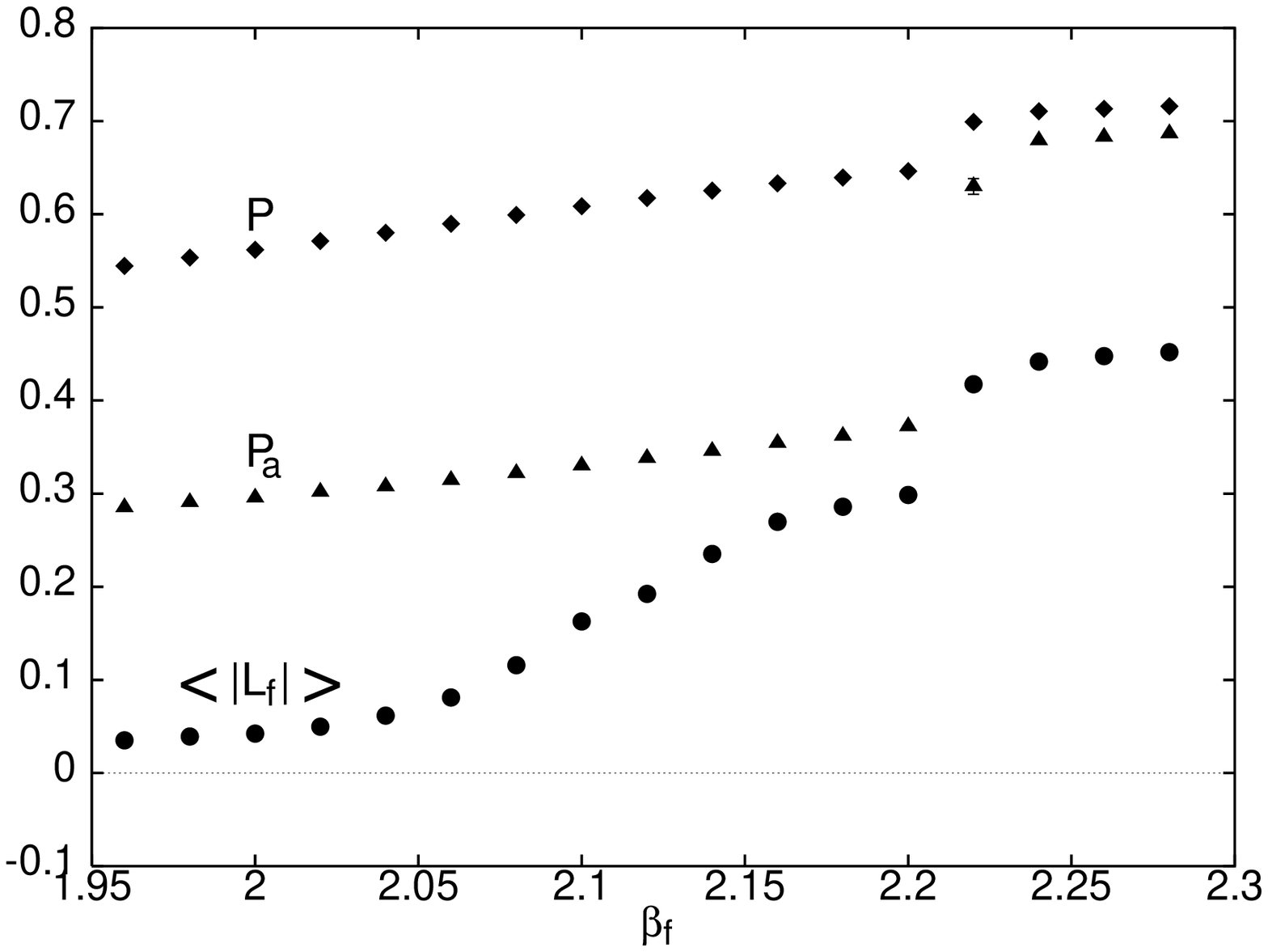}
\epsfig{height=8cm,width=8cm,file=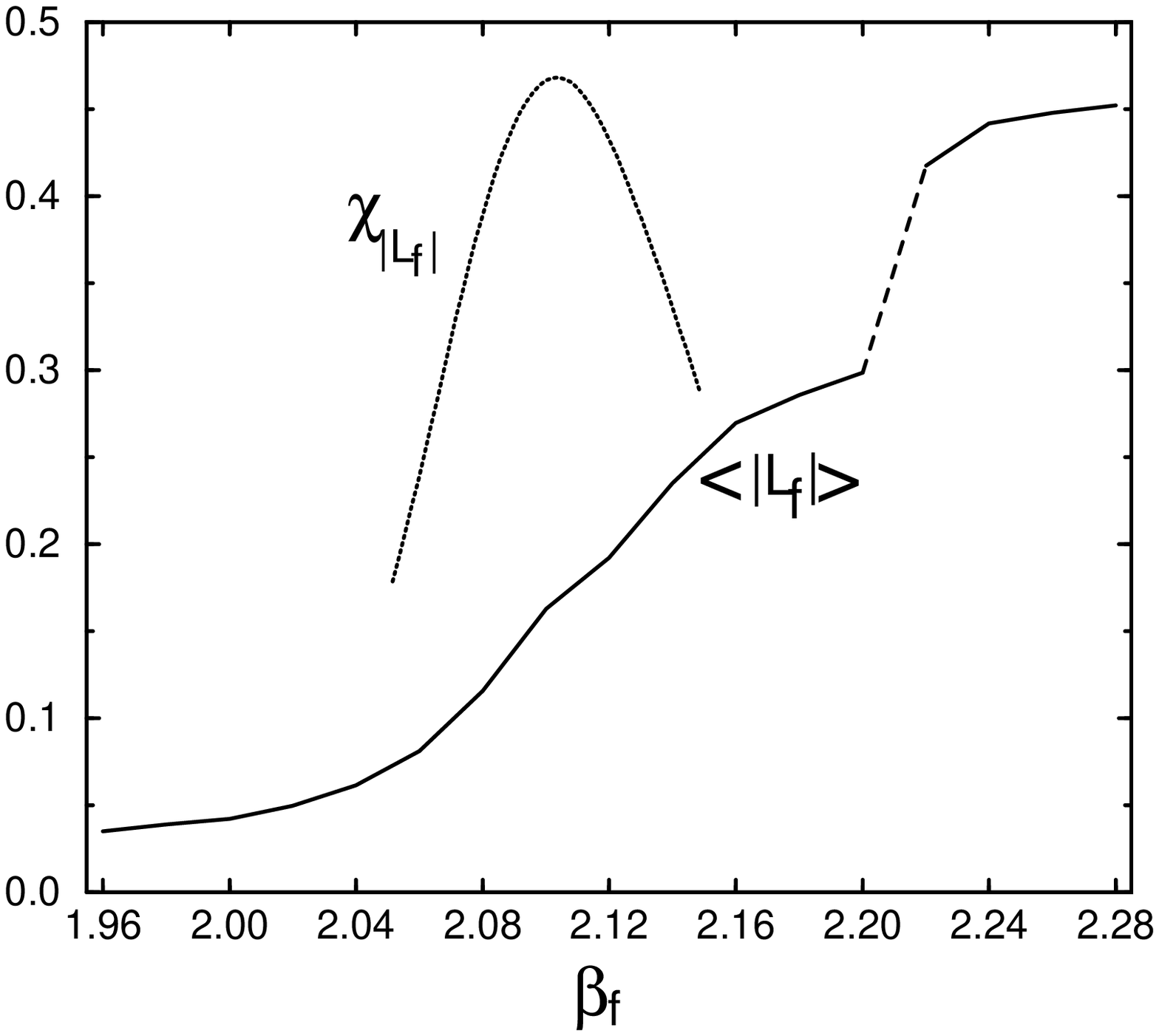}
\caption{(a) Results of a hysteresis run at $\bv$ = 0.7. 
The diamonds, triangles and circles denote the variables
$P$, $P_a$ and $\llf$ respectively. $\llf$ is seen to
rise at $\f \sim 2.1$, and a discontinuous transition is seen at $\f
\sim 2.2$.(b) $\llf$ and its susceptibility as a function of $\f$ 
at $\bv$ = 0.7, showing clearly a second order deconfinement 
transition at $\f \sim 2.1$ and a first order transition at $\f \sim 2.2$.}
\label{fg.twotr}\end{center}\end{figure} 

\begin{figure}[htbp]\begin{center}
\epsfig{height=8cm,width=12cm,file=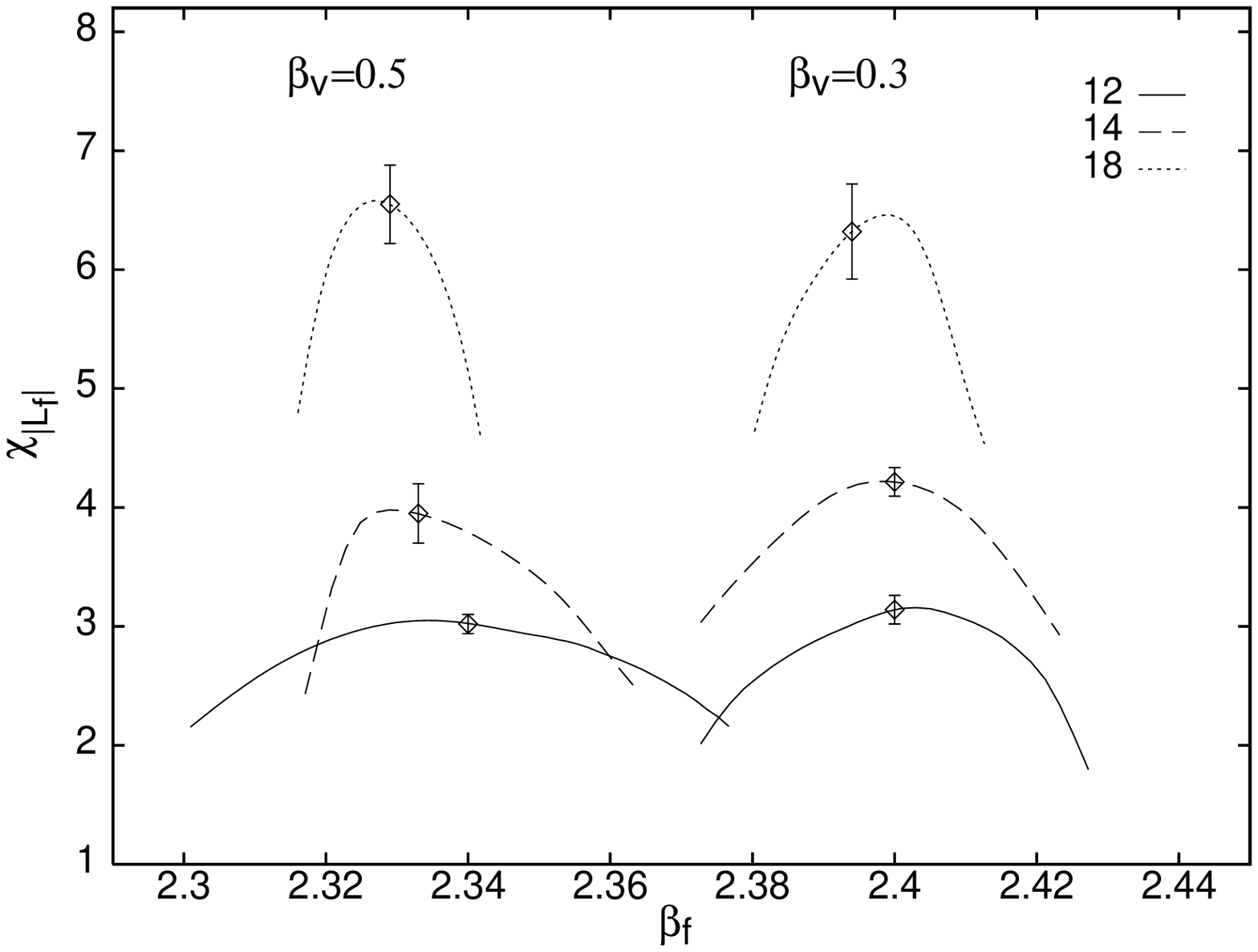}
\caption{$\clf$ as a function of $\f$ for $\nt$ = 6 lattices
with $\ns$ = 12, 14 and 18, for the action (\protect\ref{eq.chsmono}) 
at $\bv$ = 0.5 and 0.3. The values obtained directly from the long 
simulation runs are also shown with error bars. The lines were
obtained using Ferrenberg-Swendsen extrapolation.}
\label{fg.ftmono6}\end{center}\end{figure}

\begin{figure}[htbp]\begin{center}
\epsfig{height=8cm,width=12cm,file=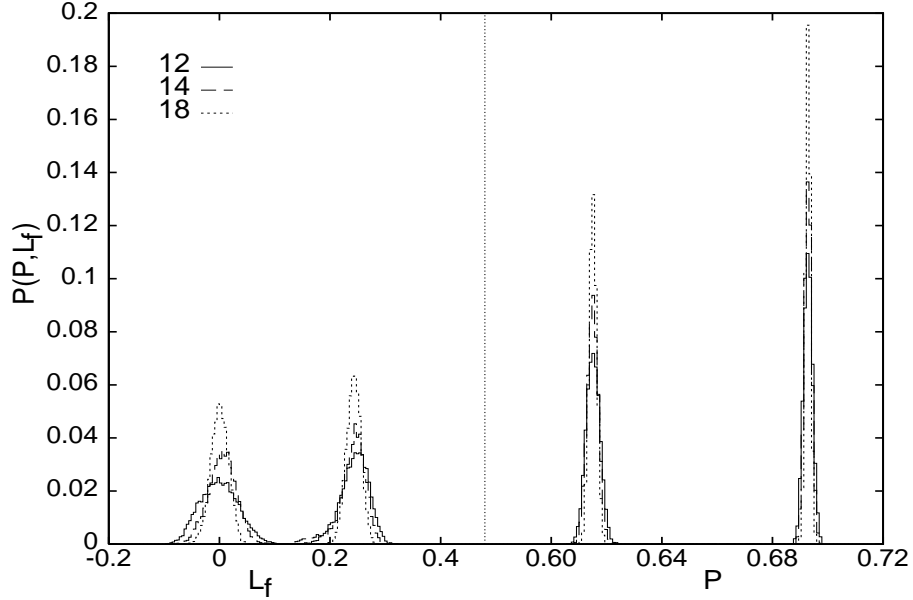}
\caption{Probability distributions of $|L_f|$ and the plaquette variable
at $\bv$ = 0.7, $\f$ = 2.12 for $\nt$ = 6 lattices with $\ns$ = 12, 14
and 18, showing metastable states indicating a first order transition.}
\label{fg.hist6}\end{center}\end{figure}

\begin{figure}[htbp]\begin{center}
\epsfig{height=8cm,width=12cm,file=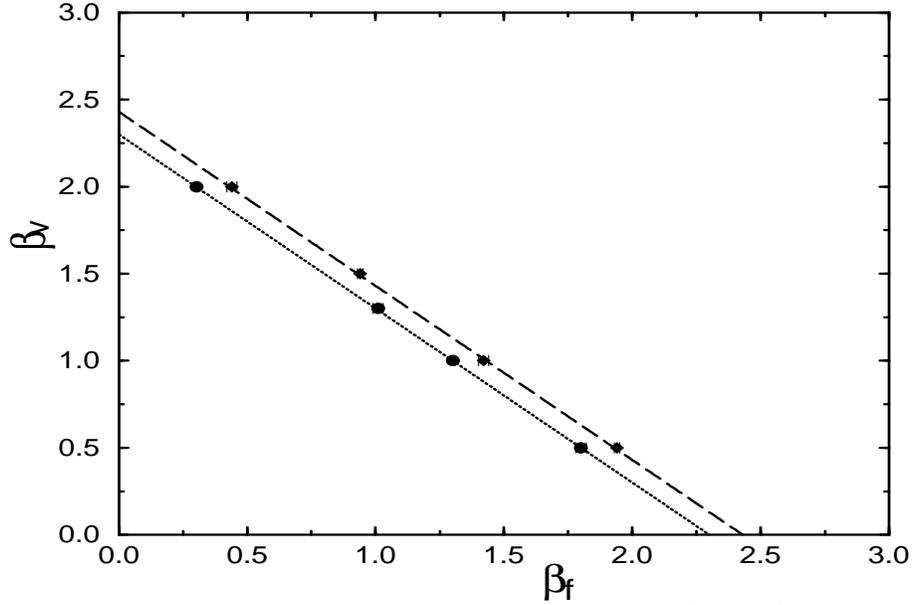}
\caption{Locations of the deconfinement transition points in the ($\f$
, $\bv$) plane. The  circles and diamonds show the simulation points
for $\nt$ = 4 and 6 lattices, respectively. The lines correspond to
 Eq. (\protect\ref{eq.dcpln}).}
\label{fg.dcpln}\end{center}\end{figure}

\newpage

\begin{figure}[htbp]\begin{center}
\epsfig{height=8cm,width=12cm,file=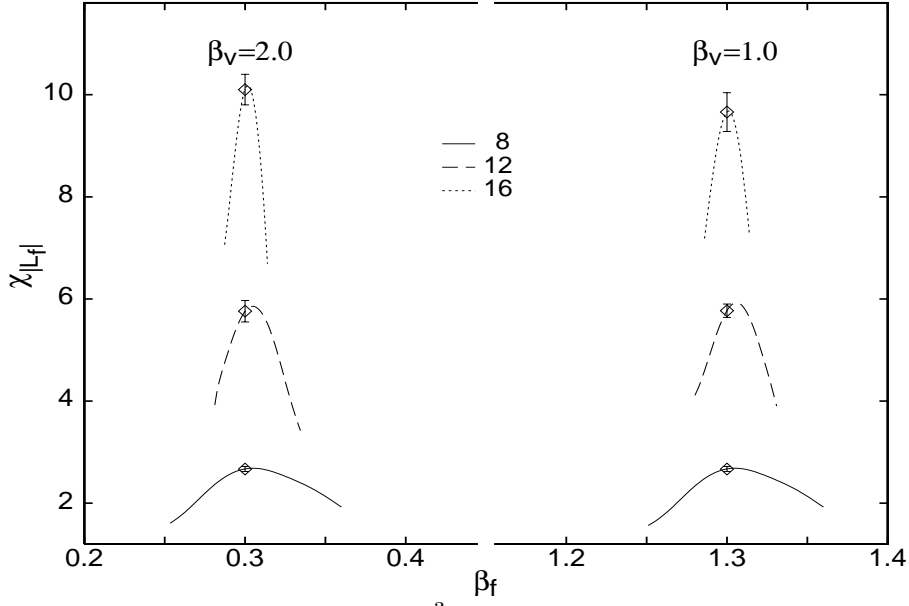}
\caption{Polyakov loop susceptibilities for $\ns^3 \times 4$ lattices
for $\ns$ = 8, 12 and 16, for $\bv$ = 1 and 2.}
\label{fg.polsusc4}\end{center}\end{figure}

\begin{figure}[htbp]\begin{center}
\epsfig{height=8cm,width=12cm,file=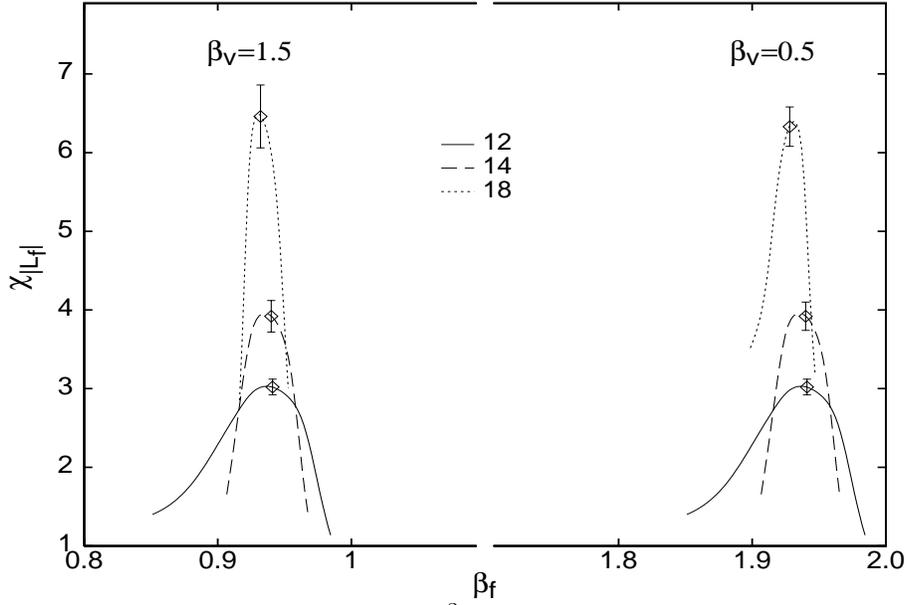}
\caption{Polyakov loop susceptibilities for $\ns^3 \times 6$ lattices
for $\ns$ = 12, 14, 18, for $\bv$ = 0.5 and 1.5. The errors are shown
for the simulation points.}
\label{fg.polsusc6}\end{center}\end{figure}

\newpage
\begin{table}[hbt]
\mediumtext
\caption[dummy]{Transition points and discontinuities for the 
first order transitions in the monopole suppressed mixed action.}
\label{tbl.bulkmono}
\begin{center}
\begin{tabular}{|c|c|c|c|c|c|}
$\beta_{vc}$ & $\beta_{fc}$ & $\Delta P$ & $\Delta P_a$ & $\Delta \llf$ & 
$|L_f|_-$ \\ 
\hline
5.0 & 0.52(2) & 0.36(2) & 0.007(3) & 0.60(4) & 0.04(3) \\ \hline
3.0 & 0.60(2) & 0.33(2) & 0.022(4) & 0.46(4) & 0.03(2) \\ \hline
2.0 & 0.76(2) & 0.26(1) & 0.08(1) & 0.38(6) & 0.07(5) \\ \hline
1.0 & 1.60(4) & 0.17(4) & 0.22(5) & 0.35(6) & 0.06(4) \\ \hline 
0.74(2) & 2.0 & 0.11(1) & 0.32(1) & 0.33(6) & 0.05(4) \\ \hline
0.59(2) & 3.0 & 0.023(4) & 0.33(2) & 0.04(5) & 0.51(4) \\ \hline
0.52(2) & 5.0 & 0.007(2) & 0.36(2) & 0.01(6) & 0.69(4) \\
\end{tabular}\end{center}
\end{table}

\begin{table}[hbt]
\mediumtext
\caption[dummy]{Values of the critical index $\omega$ 
for the monopole suppressed action.}
\label{tbl.ftmono}
\begin{center}
\begin{tabular}{|c|c|c|c|c|c|}
&\multicolumn{3}{c|}{$\nt = 4$} & \multicolumn{2}{c|}{$\nt = 6$} \\
\hline
$\bv$ & $0.3$ & $0.5$ & $0.7$ & $0.3$ & $0.5$ \\
\tableline
$\omega$ & $1.92\pm0.05$ & $1.87\pm0.05$ & $1.91\pm0.02$ 
& $1.82\pm0.08$ & $1.89\pm0.07$ \\
\end{tabular}\end{center}
\end{table}

\begin{table}[hbt]
\narrowtext
\caption[dummy]{Values of the critical index $\omega$ 
for the action with both monopoles and vortices
suppressed.}
\label{tbl.nobulk}
\begin{center}
\begin{tabular}{|c|c|c|c|c|}
& \multicolumn{2}{c|}{$\nt = 4$} & \multicolumn{2}{c|}{$\nt = 6$} \\
\hline
$\bv$ & $1.0$ & $2.0$ & $0.5$ & $1.5$ \\
\tableline
$\omega$ & $1.88\pm0.05$ & $1.93\pm0.05$ & $1.84\pm0.06$ & $1.86\pm0.07$ \\
\end{tabular}\end{center}
\end{table}


\begin{references}
\bibitem{f1} In Ref. \cite{stephenson}, a slight shift in the 
deconfinement transition line was obtained near the end of the
bulk line in going from $\nt$ = 4 to 8, unlike in Ref. \cite{manu}.
While this shift was interpreted in Ref. \cite{stephenson} as
indicating a decoupling of bulk and deconfinement effects at $\nt$ = 8,
in our opinion, it was too small and is more likely to be due to finite
size effects. In any case, the $\nt$ = 8 deconfinement line was also seen to 
bend towards, and finally merge with, the bulk line. Clearly, there were
substantial interference effects of bulk and deconfinement phenomena
there too.
\end{references}
\end{document}